\documentclass[aps,prd,onecolumn,nofootinbib,superscriptaddress,11pt]{revtex4-2}

\usepackage[T1]{fontenc}
\usepackage[utf8]{inputenc}
\usepackage{amsmath,amssymb,amsfonts,bm}
\usepackage{graphicx}
\usepackage{booktabs}
\usepackage{xcolor}

\newcommand{\dd}{\mathrm{d}}
\newcommand{\ee}{\mathrm{e}}

\newcommand{\Geff}{G_{\rm eff}}
\newcommand{\Gbar}{\bar G_{\rm eff}}

\newcommand{\E}{\mathbb{E}}
\newcommand{\Var}{\mathrm{Var}}
\newcommand{\Cov}{\mathrm{Cov}}

\newcommand{\placeholderfigure}[1]{%
\begin{center}
\fbox{%
\begin{minipage}{0.85\linewidth}
\centering
\vspace{0.35cm}
\vspace{0.35cm}
\end{minipage}}
\end{center}}

\begin{document}

\title{A Stochastic Framework for the Spherical Jeans Equation Motivated by Scalar-Tensor Gravity}

\author{Vel\'asquez-Toribio, A. M.}
\affiliation{Center for Astrophysics and Cosmology,\\ Federal University of Espírito Santo, 29075-910 Vitória - ES, Brazil}
\email{alan.toribio@ufes.br}

\date{\today}

\begin{abstract}
We develop a stochastic framework for the stationary spherical Jeans
equation, motivated by the field-dependent nature of the gravitational 
coupling in scalar--tensor theories. We model unresolved spatial fluctuations of the scalar sector 
as an effective stochastic contribution to the gravitational coupling,
$\Geff(r,\omega)=\Gbar(r)+\Gamma_G(r)\xi(r,\omega)$. This approach induces a
linear It\^o stochastic differential equation for the radial velocity dispersion
$y(r)=\sigma_r^2(r)$, defining a nonautonomous radial random flow rather than a
time-evolution problem. We derive the associated Fokker--Planck equation and obtain
integral expressions for the mean, variance, and covariance of the
radial velocity dispersion. Because the noise is additive, the
deterministic Jeans solution is recovered as the mean profile, while the
stochastic sector produces a probability band around it. We specialize
the construction to Navarro--Frenk--White, Hernquist, and Einasto halo
models and propagate the radial covariance to the projected
line-of-sight velocity dispersion. This provides a semi-analytical
framework for assessing how effective gravitational fluctuations can affect halo
kinematic observables in the stationary Jeans regime.
\end{abstract}

\maketitle

\section{Introduction}
\label{sec:introduction}

The spherical Jeans equation is one of the standard tools used to describe the equilibrium dynamics of collisionless self-gravitating systems \cite{Jeans1915}. It relates the radial velocity dispersion of a tracer population to its density profile, velocity anisotropy, and the gravitational potential. In its usual form, the equation assumes a stationary system, spherical symmetry, and no mean radial streaming motion. These assumptions make it especially useful for the internal regions of relaxed dark-matter halos, elliptical galaxies, galaxy clusters, and dwarf spheroidal galaxies \cite{BinneyTremaine}.

In the standard application of the Jeans equation, the gravitational coupling is treated as a fixed constant. However, in scalar--tensor theories of gravity the effective gravitational strength is controlled by a scalar field \cite{BransDicke1961,Clifton2012,DeFeliceTsujikawa2010}. If this scalar field has spatial fluctuations, environmental dependence, or unresolved small-scale variations, the effective gravitational coupling entering the dynamical equation may also fluctuate. The purpose of this work is to formulate a controlled stochastic version of the spherical Jeans equation in which these fluctuations are propagated into the velocity-dispersion profile.

The motivation is twofold. First, scalar--tensor gravity provides a natural setting in which the gravitational coupling is not a rigid parameter but a field-dependent quantity. Second, halo dynamics is often analyzed through deterministic Jeans models even though real systems are affected by substructure, environmental fields, departures from exact equilibrium, and possible modifications of gravity. A stochastic Jeans equation offers a simple way to quantify how such fluctuations broaden the theoretical prediction for the radial and projected velocity dispersions \cite{Arnold1998, Oksendal, Kolmogoroff1931, Ito1944, Khasminskii2012, Kunita1990, CrauelFlandoli1994, CrauelDebusscheFlandoli1997, KloedenRasmussen2011, Chandrasekhar1943Stochastic, Kandrup1980, LokasMamon2003, MamonLokas2005, MamonBivianoBoue2013, LombriserSchmidtBaldauf2012, LombriserKoyamaZhaoLi2012, EscuderoManada2022}.

We restrict the present paper to the stationary spherical Jeans regime. This is intentional. The extension to cluster outskirts, where mean radial infall, splashback physics, and the turnaround region become important, requires a generalized Jeans equation with nonzero radial streaming velocity \cite{Falco2013, MoreDiemerKravtsov2015, BarOrFouvryTremaine2021, ChiangOstrikerSchive2023, YangChiangSuSchive2024,Dalal2021, DuttaChowdhury2023, Luu2024}. Such a generalization is physically interesting, but it belongs to a different dynamical regime. Here we first construct the simplest analytical framework: a stochastic Jeans equation for the quasi-stationary halo region, together with explicit results for standard halo profiles.

The structure of the paper is as follows. In Sec.~\ref{sec:stochastic_jeans} we introduce the scalar--tensor motivation for a stochastic gravitational coupling and derive the stochastic spherical Jeans equation. In Sec.~\ref{sec:fokker_planck} we obtain the associated Fokker--Planck equation and the probability distribution of the radial velocity dispersion. In Sec.~\ref{sec:halo_models} we specialize the construction to the Navarro--Frenk--White, Hernquist, and Einasto profiles \cite{NavarroFrenkWhite1996, NavarroFrenkWhite1997, Hernquist1990, Einasto1965, Merritt2006}. In Sec. V we present numerical illustrations of the formalism and compare the analytical moment equations with Monte Carlo realizations. Also, Sec.~\ref{sec:conclusions} summarizes the results and discusses possible extensions to the infall and splashback regions.

\section{Stochastic Jeans Equation and Radial Random Flow}
\label{sec:stochastic_jeans}

\subsection{Scalar--tensor motivation for a stochastic gravitational coupling}
\label{subsec:scalar_tensor_motivation}

The motivation for introducing a stochastic gravitational coupling in the
Jeans equation comes from a basic property of scalar--tensor gravity: the
strength of the gravitational interaction is not necessarily described by
a rigid universal constant. Instead, it is controlled by the configuration
of an additional scalar degree of freedom. In a broad class of
Jordan-frame scalar--tensor theories, the gravitational sector may be
written schematically as
\begin{eqnarray}
S
&=&
\frac{1}{16\pi G_*}
\int \dd^4x\sqrt{-g}
\left[
F(\phi)R
-
Z(\phi)g^{\mu\nu}\nabla_\mu\phi\nabla_\nu\phi
-
2U(\phi)
\right]
+
S_m[g_{\mu\nu},\Psi_m].
\label{eq:scalar_tensor_action}
\end{eqnarray}
Here $G_*$ denotes a bare gravitational constant, $g_{\mu\nu}$ is the
Jordan-frame metric, $R$ is the corresponding Ricci scalar, and $\phi$ is
the scalar field. The function $F(\phi)$ controls the nonminimal coupling
between the scalar field and curvature, $Z(\phi)$ fixes the normalization
of the scalar kinetic term, and $U(\phi)$ is the scalar self-interaction
potential. The matter action $S_m[g_{\mu\nu},\Psi_m]$ depends on the
matter fields $\Psi_m$ and on the Jordan-frame metric. Therefore,
nonrelativistic matter responds to an effective gravitational strength
that depends on the scalar configuration.

In the weak-field and quasi-static regime relevant for halo dynamics, the
gravitational force entering the Jeans equation can be expressed in terms
of an effective gravitational coupling,
\begin{eqnarray}
\Geff
=
\Geff(\phi,\nabla\phi,\ldots).
\label{eq:geff_general}
\end{eqnarray}
The dots indicate that, depending on the underlying scalar--tensor model
and on the approximation scheme, $\Geff$ may depend not only on the local
value of the scalar field but also on its gradients or on quantities
associated with screening mechanisms. Thus, even if the Jeans equation is
written in an apparently Newtonian form, the coupling multiplying the
mass source may carry information about the scalar sector.

We now separate the scalar field into a smooth resolved component and an
unresolved fluctuating part,
\begin{eqnarray}
\phi(r,\omega)
&=&
\bar\phi(r)+\delta\phi(r,\omega).
\label{eq:scalar_fluctuation}
\end{eqnarray}
Here $r$ is the radial coordinate inside the spherical system and
$\omega$ labels a realization of the stochastic scalar configuration.
The field $\bar\phi(r)$ represents the coarse-grained scalar profile,
whereas $\delta\phi(r,\omega)$ represents fluctuations around this
profile. These fluctuations may effectively encode unresolved
small-scale structure, environmental variations, residual departures from
perfect equilibrium, or stochastic fluctuations of the scalar sector
itself.

Expanding the effective gravitational coupling around the smooth scalar
configuration gives, at leading order,
\begin{eqnarray}
\Geff(r,\omega)
&=&
\Geff[\bar\phi(r)]
+
\left.
\frac{\partial \Geff}{\partial \phi}
\right|_{\bar\phi}
\delta\phi(r,\omega)
+
\cdots .
\label{eq:geff_expansion_local}
\end{eqnarray}
If gradient contributions are relevant, the linear fluctuation may also
contain terms of the form
\begin{eqnarray}
\delta \Geff(r,\omega)
&=&
\left.
\frac{\partial \Geff}{\partial \phi}
\right|_{\bar\phi}
\delta\phi(r,\omega)
+
\left.
\frac{\partial \Geff}{\partial(\nabla_i\phi)}
\right|_{\bar\phi}
\nabla_i\delta\phi(r,\omega)
+
\cdots
\label{eq:geff_expansion_gradient}
\end{eqnarray}
For the present phenomenological construction, we do not need to specify
the microscopic origin of each contribution. We only assume that, after
coarse graining, the unresolved scalar-sector fluctuations induce an
effective fluctuation in the gravitational coupling. We therefore write
\begin{eqnarray}
\Geff(r,\omega)
&=&
\Gbar(r)+\delta G(r,\omega),
\label{eq:geff_split}
\end{eqnarray}
where
\begin{eqnarray}
\Gbar(r)
&\equiv&
\left\langle \Geff(r,\omega)\right\rangle
\label{eq:Gbar_definition}
\end{eqnarray}
is the ensemble-averaged, or coarse-grained, effective gravitational
coupling. The quantity $\delta G(r,\omega)$ denotes the stochastic
deviation from this mean and is assumed to satisfy
\begin{eqnarray}
\left\langle \delta G(r,\omega)\right\rangle
&=&
0.
\label{eq:deltaG_zero_mean}
\end{eqnarray}
Thus $\Gbar(r)$ determines the deterministic Jeans source, whereas
$\delta G(r,\omega)$ describes the unresolved gravitational fluctuation
around that mean.

A convenient local representation of this fluctuation is
\begin{eqnarray}
\delta G(r,\omega)
&=&
\Gamma_G(r)\xi(r,\omega),
\label{eq:deltaG_noise}
\end{eqnarray}
so that
\begin{eqnarray}
\Geff(r,\omega)
&=&
\Gbar(r)+\Gamma_G(r)\xi(r,\omega).
\label{eq:geff_stochastic}
\end{eqnarray}
In this expression, $\Gamma_G(r)$ controls the radial amplitude of the
fluctuation, while $\xi(r,\omega)$ is a zero-mean stochastic process. The
function $\Gamma_G(r)$ measures how efficiently unresolved scalar-field
fluctuations are converted into fluctuations of the gravitational
coupling. Schematically, in a perturbative regime one may regard
\begin{eqnarray}
\Gamma_G(r)\xi(r,\omega)
&\sim&
\left.
\frac{\partial \Geff}{\partial \phi}
\right|_{\bar\phi}
\delta\phi(r,\omega)
+
\cdots ,
\label{eq:Gamma_interpretation}
\end{eqnarray}
where the ellipsis may include derivative contributions, screening
effects, or other coarse-grained corrections.

In the ideal white-noise limit, the stochastic process is represented as
\begin{eqnarray}
\xi(r,\omega)
&=&
\frac{\dd W_r}{\dd r},
\label{eq:white_noise}
\end{eqnarray}
where $W_r$ is a Wiener process parametrized by the radial coordinate.
This expression is only formal: $\dd W_r/\dd r$ is not an ordinary
function, but a distributional noise. The corresponding stochastic Jeans
equation must therefore be interpreted in the It\^o sense. With this
normalization, one formally has
\begin{eqnarray}
\left\langle \xi(r)\right\rangle
&=&
0,
\qquad
\left\langle \xi(r)\xi(r')\right\rangle
=
\delta(r-r').
\label{eq:white_noise_correlation}
\end{eqnarray}
Equivalently, the gravitational fluctuation satisfies
\begin{eqnarray}
\left\langle \delta G(r)\delta G(r')\right\rangle
&=&
\Gamma_G(r)\Gamma_G(r')\delta(r-r')
\label{eq:deltaG_white_correlation}
\end{eqnarray}
in the white-noise approximation.

A more physical description would use a colored stochastic process with a
finite correlation length $\ell$,
\begin{eqnarray}
\left\langle \delta G(r)\delta G(r')\right\rangle
&=&
\Gamma_G(r)\Gamma_G(r')C_\ell(r,r'),
\label{eq:colored_noise}
\end{eqnarray}
where $C_\ell(r,r')$ is a correlation kernel. The white-noise model is
then recovered when $C_\ell(r,r')$ becomes sharply peaked around
$r=r'$. In this sense, the white-noise limit should be understood as an
idealized coarse-grained approximation, useful because it leads to a
local Fokker--Planck equation and to closed expressions for the mean,
variance, and covariance of the Jeans solution.

It is also useful to state the perturbative regime in which the
decomposition is physically meaningful. Since $\Geff$ represents the
strength of the attractive gravitational interaction, one should require
the stochastic fluctuations to remain small compared with the mean
coupling, namely
\begin{eqnarray}
|\delta G(r,\omega)|
&\ll&
\Gbar(r),
\label{eq:perturbative_condition}
\end{eqnarray}
or, more precisely, that the probability of obtaining negative or
unphysically large values of $\Geff$ is negligible. This assumption
ensures that the stochastic model describes small fluctuations around a
well-defined mean gravitational interaction.

Thus, if the radial coordinate carries dimensions, the Wiener
increment $\dd W_r$ has dimensions of $\sqrt{\rm length}$. Therefore, in
the white-noise representation, the amplitude $\Gamma_G(r)$ carries the
appropriate units required for $\Gamma_G(r)\dd W_r/\dd r$ to have the
same dimensions as a gravitational coupling. If one instead uses the
dimensionless coordinate $x=r/r_s$, the corresponding normalization
factor can be absorbed into a redefined stochastic amplitude. This is the
convention adopted below when the stochastic Jeans equation is written in
dimensionless form.
\subsection{Deterministic spherical Jeans equation}
\label{subsec:deterministic_jeans}

We consider a spherically symmetric collisionless system and define
\begin{eqnarray}
y(r)\equiv \sigma_r^2(r),
\label{eq:y_definition}
\end{eqnarray}
where $\sigma_r(r)$ is the radial velocity dispersion. The stationary spherical Jeans equation is
\begin{eqnarray}
\frac{\dd y}{\dd r}
+
\left(
\frac{\dd\ln\nu}{\dd r}
+
\frac{2\beta}{r}
\right)y
&=&
-
\Geff(r)
\frac{M(r)}{r^2},
\label{eq:deterministic_jeans}
\end{eqnarray}
where $\nu(r)$ is the tracer density, $M(r)$ is the enclosed gravitating mass, and
\begin{eqnarray}
\beta(r)
&=&
1-
\frac{\sigma_t^2(r)}{2\sigma_r^2(r)}
\label{eq:anisotropy}
\end{eqnarray}
is the orbital anisotropy parameter. For compactness we define
\begin{eqnarray}
F(r)
&=&
\frac{\dd\ln\nu}{\dd r}
+
\frac{2\beta(r)}{r},
\qquad
H(r)
=
\frac{M(r)}{r^2}.
\label{eq:FH_definitions}
\end{eqnarray}
Then Eq.~\eqref{eq:deterministic_jeans} becomes
\begin{eqnarray}
\frac{\dd y}{\dd r}
&=&
-F(r)y-\Geff(r)H(r).
\label{eq:jeans_compact}
\end{eqnarray}

\subsection{Stochastic Jeans equation}
\label{subsec:stochastic_jeans_equation}

Using Eq.~\eqref{eq:geff_stochastic}, the Jeans equation becomes a stochastic differential equation along the radial coordinate,
\begin{eqnarray}
\dd y
&=&
\left[-F(r)y-\Gbar(r)H(r)\right]\dd r
-
B(r)\dd W_r,
\label{eq:stochastic_jeans_sde}
\end{eqnarray}
where
\begin{eqnarray}
B(r)
&=&
\Gamma_G(r)H(r).
\label{eq:B_definition}
\end{eqnarray}
The stochastic term is additive because the diffusion coefficient does not depend on the state variable $y$. This property makes the model analytically tractable and ensures that, for deterministic boundary data and Gaussian noise, the velocity-dispersion profile is Gaussian distributed.

Equation~\eqref{eq:stochastic_jeans_sde} defines a radial random flow. The radial coordinate plays the role of an evolution parameter, while the stochastic process encodes unresolved scalar-field fluctuations. Since $F(r)$, $H(r)$, and $B(r)$ are explicitly radial-dependent, the corresponding stochastic flow is nonautonomous. Therefore, one should not automatically interpret its probability distribution as a stationary invariant measure. A more appropriate interpretation is that Eq.~\eqref{eq:stochastic_jeans_sde} defines a random radial boundary-value problem, or equivalently a pullback distribution determined by the outer boundary condition.

Let
\begin{eqnarray}
I(r)
&=&
\exp\left[
\int^r F(u)\dd u
\right]
\label{eq:integrating_factor_general}
\end{eqnarray}
be the deterministic integrating factor. Imposing an outer boundary condition at $r=r_{\rm out}$, the formal solution is
\begin{eqnarray}
y(r)
&=&
\frac{I(r_{\rm out})}{I(r)}y(r_{\rm out})
+
\frac{1}{I(r)}
\int_r^{r_{\rm out}} I(u)\Gbar(u)H(u)\dd u
+
\frac{1}{I(r)}
\int_r^{r_{\rm out}} I(u)B(u)\dd W_u.
\label{eq:formal_solution_general}
\end{eqnarray}
For the common boundary condition $y(r_{\rm out})=0$, the first term vanishes.

\section{Fokker--Planck equation and radial probability distribution}
\label{sec:fokker_planck}

The stochastic Jeans equation derived in the previous section defines a
one-dimensional nonautonomous It\^o diffusion for the radial
velocity-dispersion variable
\begin{eqnarray}
Y_r \equiv y(r)=\sigma_r^2(r).
\end{eqnarray}
The radial coordinate is not a physical time. Nevertheless, it plays the
role of an ordering parameter for the radial boundary-value problem.
Accordingly, the probability density $P(y,r)$ should be interpreted as a
radial probability distribution, describing how uncertainties induced by
the stochastic gravitational sector are propagated along the halo
profile.

The drift and diffusion coefficients associated with
Eq.~\eqref{eq:stochastic_jeans_sde} are
\begin{eqnarray}
a(r,y)
&=&
-F(r)y-\Gbar(r)H(r),
\qquad
b(r)
=
-B(r).
\label{eq:drift_diffusion_coefficients}
\end{eqnarray}
Since the diffusion coefficient is independent of $y$, the stochastic
term is additive. The forward Kolmogorov equation, usually referred to in
physics as the Fokker--Planck equation, then gives
\begin{eqnarray}
\frac{\partial P}{\partial r}
&=&
\frac{\partial}{\partial y}
\left[
\left(F(r)y+\Gbar(r)H(r)\right)P
\right]
+
\frac{1}{2}B^2(r)
\frac{\partial^2P}{\partial y^2}.
\label{eq:fokker_planck}
\end{eqnarray}
The first term represents the deterministic Jeans drift, while the
second term describes diffusion in the variable $y=\sigma_r^2$ generated
by the stochastic contribution $\Gamma_G\xi$ to the effective
gravitational coupling.

The local form of Eq.~\eqref{eq:fokker_planck} relies on the
white-noise coarse-grained limit, $\xi(r)=\dd W_r/\dd r$. If the
fluctuations of the gravitational coupling are modeled as a colored
process with finite correlation length, the dynamics is no longer
Markovian in $y$ alone. In that case, the Fokker--Planck description must
either be generalized to a nonlocal evolution equation or reformulated in
an enlarged state space including the stochastic forcing as an auxiliary
variable. In the following, we keep the white-noise limit as the minimal
analytically tractable model.

We now use the formal solution obtained in
Eq.~\eqref{eq:formal_solution_general} to compute the statistical
moments of the velocity-dispersion profile. Taking the expectation value,
and using the fact that the It\^o integral has zero mean for an adapted
square-integrable integrand, one finds
\begin{eqnarray}
\E[Y_r]
&=&
\frac{I(r_{\rm out})}{I(r)}Y_{r_{\rm out}}
+
\frac{1}{I(r)}
\int_r^{r_{\rm out}}
I(u)\Gbar(u)H(u)\dd u.
\label{eq:mean_general}
\end{eqnarray}
Thus the stochastic mean coincides with the deterministic Jeans solution
with the same outer boundary condition. Therefore, in the present
additive-noise model, the leading stochastic effect is not a displacement
of the mean profile, but the generation of a probability band around it.

The variance is determined by the stochastic integral appearing in
Eq.~\eqref{eq:formal_solution_general}. Since the deterministic part does
not contribute to the variance, the relevant term is
\begin{eqnarray}
\mathcal{S}(r)
&=&
\frac{1}{I(r)}
\int_r^{r_{\rm out}}
I(u)B(u)\dd W_u .
\end{eqnarray}
Using the It\^o isometry,
\begin{eqnarray}
\E\left[
\left(
\int_r^{r_{\rm out}}
I(u)B(u)\dd W_u
\right)^2
\right]
&=&
\int_r^{r_{\rm out}}
I^2(u)B^2(u)\dd u,
\end{eqnarray}
we obtain
\begin{eqnarray}
\Var[Y_r]
&=&
\frac{1}{I^2(r)}
\int_r^{r_{\rm out}}
I^2(u)B^2(u)\dd u.
\label{eq:variance_general}
\end{eqnarray}
This expression shows explicitly how the fluctuation amplitude
$B(r)=\Gamma_G(r)H(r)$ controls the stochastic broadening of the radial
velocity-dispersion profile.

The same argument gives the covariance between the velocity dispersion
evaluated at two different radii. The stochastic parts of $Y_r$ and
$Y_s$ contain It\^o integrals over the intervals $[r,r_{\rm out}]$ and
$[s,r_{\rm out}]$, respectively. Their common stochastic contribution is
therefore supported on the overlap interval
$[\max(r,s),r_{\rm out}]$. Applying the It\^o isometry to this overlap
gives
\begin{eqnarray}
\Cov[Y_r,Y_s]
&=&
\frac{1}{I(r)I(s)}
\int_{\max(r,s)}^{r_{\rm out}}
I^2(u)B^2(u)\dd u.
\label{eq:covariance_general}
\end{eqnarray}
The stochastic Jeans model therefore predicts not only a mean
velocity-dispersion profile, but also a radial covariance structure. This
covariance is the object that should be propagated when constructing
projected observables such as the line-of-sight velocity dispersion.

Because the stochastic Jeans equation is linear and the noise is
Gaussian, the solution is Gaussian at each fixed radius whenever the
outer boundary value is deterministic. The local probability density can
therefore be written as
\begin{eqnarray}
P(y,r)
&=&
\frac{1}{\sqrt{2\pi\Sigma^2(r)}}
\exp\left[
-
\frac{\left(y-\mu(r)\right)^2}{2\Sigma^2(r)}
\right],
\label{eq:gaussian_distribution}
\end{eqnarray}
where
\begin{eqnarray}
\mu(r)
&=&
\E[Y_r],
\qquad
\Sigma^2(r)
=
\Var[Y_r].
\label{eq:mu_sigma_def}
\end{eqnarray}
The deterministic Jeans model is recovered in the limit
$B(r)\rightarrow 0$, or equivalently $\Gamma_G(r)\rightarrow 0$, in
which the probability distribution collapses to a delta distribution
centered on the deterministic Jeans solution.

The results obtained in this section are independent of the particular
halo profile. All model dependence enters through the functions $F(r)$,
$H(r)$, the integrating factor $I(r)$, and the stochastic amplitude
$B(r)=\Gamma_G(r)H(r)$. In the next section, these general expressions
are specialized to the Navarro--Frenk--White, Hernquist, and Einasto
profiles.

\section{Analytical Halo Models}
\label{sec:halo_models}

In this section we specialize the general stochastic Jeans construction
to three standard spherical halo profiles: the Navarro--Frenk--White
(NFW), Hernquist, and Einasto models. These examples are useful because
they represent different analytical descriptions of halo structure. The
NFW profile is the standard benchmark motivated by cosmological
$N$-body simulations, the Hernquist profile provides a finite-mass model
with the same inner cusp as NFW, and the Einasto profile introduces a
continuously varying logarithmic slope, allowing one to test the effect
of a more flexible halo shape.

We introduce the dimensionless radial coordinate
\begin{eqnarray}
x=\frac{r}{r_s},
\label{eq:x_definition}
\end{eqnarray}
where $r_s$ is the characteristic scale radius of the profile under
consideration. For the Einasto profile, we identify this scale with
$r_{-2}$, the radius at which the logarithmic slope is equal to $-2$.
For definiteness, we assume in this section that the tracer density
$\nu(r)$ has the same radial dependence as the corresponding halo
density profile. This choice keeps the expressions compact and allows us
to isolate the effect of the different halo profiles. More generally,
one may use a tracer distribution different from the gravitating mass
density. In that case, the enclosed mass $M(r)$ must still be computed
from the gravitating profile, while the logarithmic derivative entering
the Jeans coefficient must be computed from the tracer density $\nu(r)$.

For constant anisotropy parameter $\beta$, the dimensionless Jeans
equation can be written as
\begin{eqnarray}
\frac{\dd y}{\dd x}
+
\mathcal{F}(x)y
&=&
-\bar{\mathcal A}(x),
\label{eq:dimensionless_jeans}
\end{eqnarray}
where
\begin{eqnarray}
y(x)
&=&
\sigma_r^2(x),
\qquad
\mathcal{F}(x)
=
\frac{\dd\ln\nu}{\dd x}
+
\frac{2\beta}{x}.
\label{eq:F_dimensionless}
\end{eqnarray}
The deterministic source term is
\begin{eqnarray}
\bar{\mathcal A}(x)
&=&
\frac{\Gbar(x)M(x)}{r_s x^2},
\label{eq:A_dimensionless_general}
\end{eqnarray}
whereas the corresponding stochastic amplitude is
\begin{eqnarray}
\mathcal{B}(x)
&=&
\frac{\Gamma_G(x)M(x)}{r_s x^2}.
\label{eq:B_dimensionless_general}
\end{eqnarray}
Thus $\bar{\mathcal A}(x)$ is the mean gravitational source in the Jeans
equation, while $\mathcal{B}(x)$ controls the amplitude of the random
forcing generated by fluctuations of the effective gravitational
coupling. If the stochastic equation is written directly in the
dimensionless coordinate $x$, constant normalization factors associated
with the transformation of the Wiener process can be absorbed into the
definition of $\Gamma_G$.

For any given profile, the integrating factor is
\begin{eqnarray}
I(x)
&=&
\exp\left[
\int^x \mathcal{F}(u)\dd u
\right].
\label{eq:dimensionless_integrating_factor}
\end{eqnarray}
The general results of Sec.~\ref{sec:fokker_planck} then give
\begin{eqnarray}
\mu(x)
&=&
\frac{1}{I(x)}
\int_x^{x_{\rm out}}
I(u)\bar{\mathcal A}(u)\dd u,
\label{eq:mean_dimensionless_general}
\\
\Sigma^2(x)
&=&
\frac{1}{I^2(x)}
\int_x^{x_{\rm out}}
I^2(u)\mathcal{B}^2(u)\dd u.
\label{eq:variance_dimensionless_general}
\end{eqnarray}
Here $x_{\rm out}$ is the outer boundary at which the radial Jeans
problem is closed. Keeping this radius finite is physically natural,
since it may represent a virial radius, a splashback radius, an
observational aperture, or the scale beyond which the stationary Jeans
description is no longer assumed to apply. It is also mathematically
useful, especially for profiles whose mass does not converge rapidly at
large radius.

\subsection{Navarro--Frenk--White profile}
\label{subsec:nfw}

The Navarro--Frenk--White profile is defined by \cite{NavarroFrenkWhite1996}
\begin{eqnarray}
\rho_{\rm NFW}(x)
&=&
\frac{\rho_s}{x(1+x)^2}.
\label{eq:nfw_density}
\end{eqnarray}
It behaves as $\rho_{\rm NFW}\propto x^{-1}$ in the inner region and as
$\rho_{\rm NFW}\propto x^{-3}$ at large radius. The corresponding
enclosed mass is
\begin{eqnarray}
M_{\rm NFW}(x)
&=&
M_s m_{\rm NFW}(x),
\qquad
m_{\rm NFW}(x)
=
\ln(1+x)-\frac{x}{1+x},
\label{eq:nfw_mass}
\end{eqnarray}
where $M_s=4\pi\rho_s r_s^3$. The logarithmic term in
$m_{\rm NFW}(x)$ implies that the NFW mass grows logarithmically at
large radius. Therefore, in a Jeans analysis it is appropriate to impose
a finite outer boundary $x_{\rm out}$.

If the tracer density follows the NFW radial dependence, then
\begin{eqnarray}
\frac{\dd\ln\nu_{\rm NFW}}{\dd x}
&=&
-\frac{1}{x}
-
\frac{2}{1+x}.
\label{eq:nfw_log_slope}
\end{eqnarray}
For constant anisotropy, the Jeans coefficient becomes
\begin{eqnarray}
\mathcal{F}_{\rm NFW}(x)
&=&
\frac{2\beta-1}{x}
-
\frac{2}{1+x}.
\label{eq:nfw_F}
\end{eqnarray}
The integrating factor is therefore
\begin{eqnarray}
I_{\rm NFW}(x)
&=&
x^{2\beta-1}(1+x)^{-2},
\label{eq:nfw_integrating_factor}
\end{eqnarray}
up to an irrelevant multiplicative constant.

The deterministic and stochastic source functions are
\begin{eqnarray}
\bar{\mathcal A}_{\rm NFW}(x)
&=&
\frac{\Gbar(x)M_s}{r_s}
\frac{m_{\rm NFW}(x)}{x^2},
\label{eq:nfw_A}
\\
\mathcal{B}_{\rm NFW}(x)
&=&
\frac{\Gamma_G(x)M_s}{r_s}
\frac{m_{\rm NFW}(x)}{x^2}.
\label{eq:nfw_B}
\end{eqnarray}
Consequently, the mean radial velocity dispersion is
\begin{eqnarray}
\mu_{\rm NFW}(x)
&=&
\frac{1}{I_{\rm NFW}(x)}
\int_x^{x_{\rm out}}
I_{\rm NFW}(u)
\bar{\mathcal A}_{\rm NFW}(u)\dd u,
\label{eq:nfw_mean}
\end{eqnarray}
and the stochastic variance is
\begin{eqnarray}
\Sigma^2_{\rm NFW}(x)
&=&
\frac{1}{I_{\rm NFW}^2(x)}
\int_x^{x_{\rm out}}
I_{\rm NFW}^2(u)
\mathcal{B}_{\rm NFW}^2(u)\dd u.
\label{eq:nfw_variance}
\end{eqnarray}
These expressions are semi-analytical: the profile dependence is written
in terms of elementary functions, and the remaining computation is a
one-dimensional quadrature. For special choices of $\beta$, $\Gbar(x)$,
and $\Gamma_G(x)$, further analytical simplifications may be possible,
but the integral representation is the most transparent form for
numerical applications and comparisons among profiles.

\subsection{Hernquist profile}
\label{subsec:hernquist}

The Hernquist profile is given by \cite{Hernquist1990}
\begin{eqnarray}
\rho_{\rm H}(x)
&=&
\frac{\rho_s}{x(1+x)^3}.
\label{eq:hernquist_density}
\end{eqnarray}
It has the same inner cusp as the NFW profile,
$\rho_{\rm H}\propto x^{-1}$, but a steeper outer decline,
$\rho_{\rm H}\propto x^{-4}$. As a consequence, its total mass is finite.
With a convenient normalization, the enclosed mass is
\begin{eqnarray}
M_{\rm H}(x)
&=&
M_{\rm tot}
\frac{x^2}{(1+x)^2}.
\label{eq:hernquist_mass}
\end{eqnarray}
This finite-mass property makes the Hernquist model a useful benchmark
for testing the stochastic Jeans construction.

For a tracer density proportional to Eq.~\eqref{eq:hernquist_density},
one obtains
\begin{eqnarray}
\frac{\dd\ln\nu_{\rm H}}{\dd x}
&=&
-\frac{1}{x}
-
\frac{3}{1+x}.
\label{eq:hernquist_log_slope}
\end{eqnarray}
Hence
\begin{eqnarray}
\mathcal{F}_{\rm H}(x)
&=&
\frac{2\beta-1}{x}
-
\frac{3}{1+x}.
\label{eq:hernquist_F}
\end{eqnarray}
The corresponding integrating factor is
\begin{eqnarray}
I_{\rm H}(x)
&=&
x^{2\beta-1}(1+x)^{-3}.
\label{eq:hernquist_integrating_factor}
\end{eqnarray}

The gravitational source takes a particularly simple form because
\begin{eqnarray}
\frac{M_{\rm H}(x)}{x^2}
&=&
\frac{M_{\rm tot}}{(1+x)^2}.
\end{eqnarray}
Therefore,
\begin{eqnarray}
\bar{\mathcal A}_{\rm H}(x)
&=&
\frac{\Gbar(x)M_{\rm tot}}{r_s}
\frac{1}{(1+x)^2},
\label{eq:hernquist_A}
\\
\mathcal{B}_{\rm H}(x)
&=&
\frac{\Gamma_G(x)M_{\rm tot}}{r_s}
\frac{1}{(1+x)^2}.
\label{eq:hernquist_B}
\end{eqnarray}
The mean and variance are then
\begin{eqnarray}
\mu_{\rm H}(x)
&=&
\frac{1}{I_{\rm H}(x)}
\int_x^{x_{\rm out}}
I_{\rm H}(u)
\bar{\mathcal A}_{\rm H}(u)\dd u,
\label{eq:hernquist_mean}
\\
\Sigma^2_{\rm H}(x)
&=&
\frac{1}{I_{\rm H}^2(x)}
\int_x^{x_{\rm out}}
I_{\rm H}^2(u)
\mathcal{B}_{\rm H}^2(u)\dd u.
\label{eq:hernquist_variance}
\end{eqnarray}
Compared with the NFW case, the Hernquist model has a simpler
gravitational source and better asymptotic behavior. For constant
$\Gbar$ and $\Gamma_G$, the kernels entering the mean and variance are
rational functions multiplied by the integrating factor. This makes the
Hernquist profile especially useful for checking numerical
implementations of the stochastic Jeans formalism.

\subsection{Einasto profile}
\label{subsec:einasto}

The Einasto profile is written as \cite{Einasto1965}
\begin{eqnarray}
\rho_{\rm E}(x)
&=&
\rho_{-2}
\exp\left[
-\frac{2}{\alpha_E}
\left(x^{\alpha_E}-1\right)
\right],
\label{eq:einasto_density}
\end{eqnarray}
where
\begin{eqnarray}
x=\frac{r}{r_{-2}}.
\end{eqnarray}
Here $r_{-2}$ is the radius at which the logarithmic slope is equal to
$-2$, and $\alpha_E$ controls the curvature of the profile. Unlike the
NFW and Hernquist models, the Einasto profile does not have a fixed
power-law slope. Instead,
\begin{eqnarray}
\frac{\dd\ln\rho_{\rm E}}{\dd\ln x}
&=&
-2x^{\alpha_E}.
\label{eq:einasto_log_slope_ln}
\end{eqnarray}
Thus the logarithmic slope varies continuously with radius, becoming
progressively steeper toward the outer halo.

If the tracer density follows the Einasto radial dependence, then
\begin{eqnarray}
\frac{\dd\ln\nu_{\rm E}}{\dd x}
&=&
-2x^{\alpha_E-1}.
\label{eq:einasto_log_slope}
\end{eqnarray}
For constant anisotropy,
\begin{eqnarray}
\mathcal{F}_{\rm E}(x)
&=&
-2x^{\alpha_E-1}
+
\frac{2\beta}{x}.
\label{eq:einasto_F}
\end{eqnarray}
The integrating factor is
\begin{eqnarray}
I_{\rm E}(x)
&=&
x^{2\beta}
\exp\left[
-\frac{2}{\alpha_E}x^{\alpha_E}
\right],
\label{eq:einasto_integrating_factor}
\end{eqnarray}
again up to an irrelevant multiplicative constant.

The enclosed mass follows from
\begin{eqnarray}
M_{\rm E}(x)
&=&
4\pi r_{-2}^3
\int_0^x
\rho_{\rm E}(u)u^2\dd u.
\end{eqnarray}
For a general value of $\alpha_E$, this integral is not elementary. It
can be expressed in terms of the lower incomplete gamma function,
\begin{eqnarray}
\gamma(a,z)
&=&
\int_0^z t^{a-1}e^{-t}\dd t.
\label{eq:lower_incomplete_gamma_def}
\end{eqnarray}
Using this definition, the enclosed mass is
\begin{eqnarray}
M_{\rm E}(x)
&=&
4\pi\rho_{-2}r_{-2}^3
\ee^{2/\alpha_E}
\frac{1}{\alpha_E}
\left(\frac{\alpha_E}{2}\right)^{3/\alpha_E}
\gamma\left(
\frac{3}{\alpha_E},
\frac{2x^{\alpha_E}}{\alpha_E}
\right).
\label{eq:einasto_mass}
\end{eqnarray}
It is also useful to define the total Einasto mass,
\begin{eqnarray}
M_{{\rm E},\infty}
&=&
4\pi\rho_{-2}r_{-2}^3
\ee^{2/\alpha_E}
\frac{1}{\alpha_E}
\left(\frac{\alpha_E}{2}\right)^{3/\alpha_E}
\Gamma\left(\frac{3}{\alpha_E}\right),
\label{eq:einasto_total_mass}
\end{eqnarray}
where $\Gamma(a)$ is the complete gamma function. Then
\begin{eqnarray}
M_{\rm E}(x)
&=&
M_{{\rm E},\infty}\,m_{\rm E}(x),
\label{eq:einasto_mass_normalized}
\end{eqnarray}
with
\begin{eqnarray}
m_{\rm E}(x)
&=&
\frac{
\gamma\left(
3/\alpha_E,
2x^{\alpha_E}/\alpha_E
\right)
}{
\Gamma\left(3/\alpha_E\right)
}.
\label{eq:einasto_mass_fraction}
\end{eqnarray}
The function $m_{\rm E}(x)$ is the fraction of the total Einasto mass
enclosed within radius $x$.

The deterministic source and stochastic amplitude are therefore
\begin{eqnarray}
\bar{\mathcal A}_{\rm E}(x)
&=&
\frac{\Gbar(x)M_{\rm E}(x)}{r_{-2}x^2},
\label{eq:einasto_A}
\\
\mathcal{B}_{\rm E}(x)
&=&
\frac{\Gamma_G(x)M_{\rm E}(x)}{r_{-2}x^2}.
\label{eq:einasto_B}
\end{eqnarray}
The corresponding mean and variance are
\begin{eqnarray}
\mu_{\rm E}(x)
&=&
\frac{1}{I_{\rm E}(x)}
\int_x^{x_{\rm out}}
I_{\rm E}(u)
\bar{\mathcal A}_{\rm E}(u)\dd u,
\label{eq:einasto_mean}
\\
\Sigma^2_{\rm E}(x)
&=&
\frac{1}{I_{\rm E}^2(x)}
\int_x^{x_{\rm out}}
I_{\rm E}^2(u)
\mathcal{B}_{\rm E}^2(u)\dd u.
\label{eq:einasto_variance}
\end{eqnarray}
The Einasto case is therefore semi-analytical in a slightly different
sense from the NFW and Hernquist cases: the enclosed mass already
involves a special function, but once this function is specified, the
mean and variance again reduce to one-dimensional quadratures .

\subsection{Analytical structure and comparison}
\label{subsec:analytical_structure}

The three profiles lead to the same formal stochastic Jeans structure.
In each case, the mean profile is obtained from a weighted integral of
the deterministic gravitational source $\bar{\mathcal A}(x)$, while the
variance is obtained from a weighted integral of the squared stochastic
amplitude $\mathcal{B}^2(x)$. The integrating factor $I(x)$ encodes the
combined effect of the tracer logarithmic slope and the anisotropy
parameter.

For the NFW and Hernquist profiles, all functions entering the Jeans
kernel are elementary. The NFW profile contains a logarithmic mass
function and therefore requires a finite outer boundary in practical
applications. The Hernquist profile has a finite total mass and a simpler
source term, making it a useful benchmark for numerical checks. The
Einasto profile has a finite total mass and a curved logarithmic slope,
but its enclosed mass involves the lower incomplete gamma function. This
is the only special function needed in the present analytical
construction.

Once $\Gbar(x)$, $\Gamma_G(x)$, $\beta$, and $x_{\rm out}$ are specified,
Eqs.~\eqref{eq:mean_dimensionless_general} and
\eqref{eq:variance_dimensionless_general} determine the mean radial
velocity-dispersion profile and the stochastic probability band for each
halo model. These quantities provide the input for the numerical
diagnostics discussed in the next section.

\subsection{Projected line-of-sight velocity dispersion}

The radial Jeans solution is connected with observable kinematics through
the projected line-of-sight velocity dispersion. For consistency with the
finite radial boundary used to close the Jeans problem, we define the
projected tracer density as
\begin{equation}
\Sigma_{\rm tr}(R)
=
2\int_R^{r_{\rm out}}
\frac{\nu(r)r\,dr}{\sqrt{r^2-R^2}} .
\end{equation}
The deterministic projected velocity dispersion is then
\begin{equation}
\sigma_{\rm los,det}^2(R)
=
\frac{2}{\Sigma_{\rm tr}(R)}
\int_R^{r_{\rm out}}
\left[
1-\beta(r)\frac{R^2}{r^2}
\right]
\frac{\nu(r)\sigma_r^2(r)r\,dr}{\sqrt{r^2-R^2}} .
\end{equation}

In the stochastic model, the radial quantity
$Y_r=\sigma_r^2(r)$ is a random field with mean $\mu(r)$ and covariance
\begin{equation}
C_y(r,s)=\mathrm{Cov}[Y_r,Y_s] .
\end{equation}
It is useful to introduce the projection kernel
\begin{equation}
K(R,r)
=
\frac{2}{\Sigma_{\rm tr}(R)}
\left[
1-\beta(r)\frac{R^2}{r^2}
\right]
\frac{\nu(r)r}{\sqrt{r^2-R^2}} .
\end{equation}
The projected mean is therefore
\begin{equation}
\bar\sigma_{\rm los}^2(R)
=
\int_R^{r_{\rm out}} K(R,r)\mu(r)\,dr .
\end{equation}

The stochastic uncertainty is propagated through the same projection
operator. Hence, the covariance between two projected radii $R$ and $R'$ is
\begin{equation}
C_{\rm los}(R,R')
=
\int_R^{r_{\rm out}}dr
\int_{R'}^{r_{\rm out}}ds\,
K(R,r)K(R',s)C_y(r,s).
\end{equation}
In particular, the variance of the projected line-of-sight dispersion at a
fixed projected radius is
\begin{equation}
\mathrm{Var}\!\left[\sigma_{\rm los}^2(R)\right]
=
C_{\rm los}(R,R).
\end{equation}
This expression shows that the observable uncertainty bands are not inserted
phenomenologically, but follow from the radial stochastic Jeans covariance
after line-of-sight projection.

\section{Some Numerical Results}
\label{sec:numerical_results}

This section should present the main numerical illustrations of the formalism. The purpose is not to perform a full parameter inference, but to show how scalar-field fluctuations propagate into halo kinematic observables. Unless otherwise stated, one may adopt dimensionless units with $r_s=1$, $G_0=1$, and a fixed outer radius $x_{\rm out}$ sufficiently large for the chosen profile.

\begin{figure}[t]
\centering
\includegraphics[width=0.4600\linewidth]{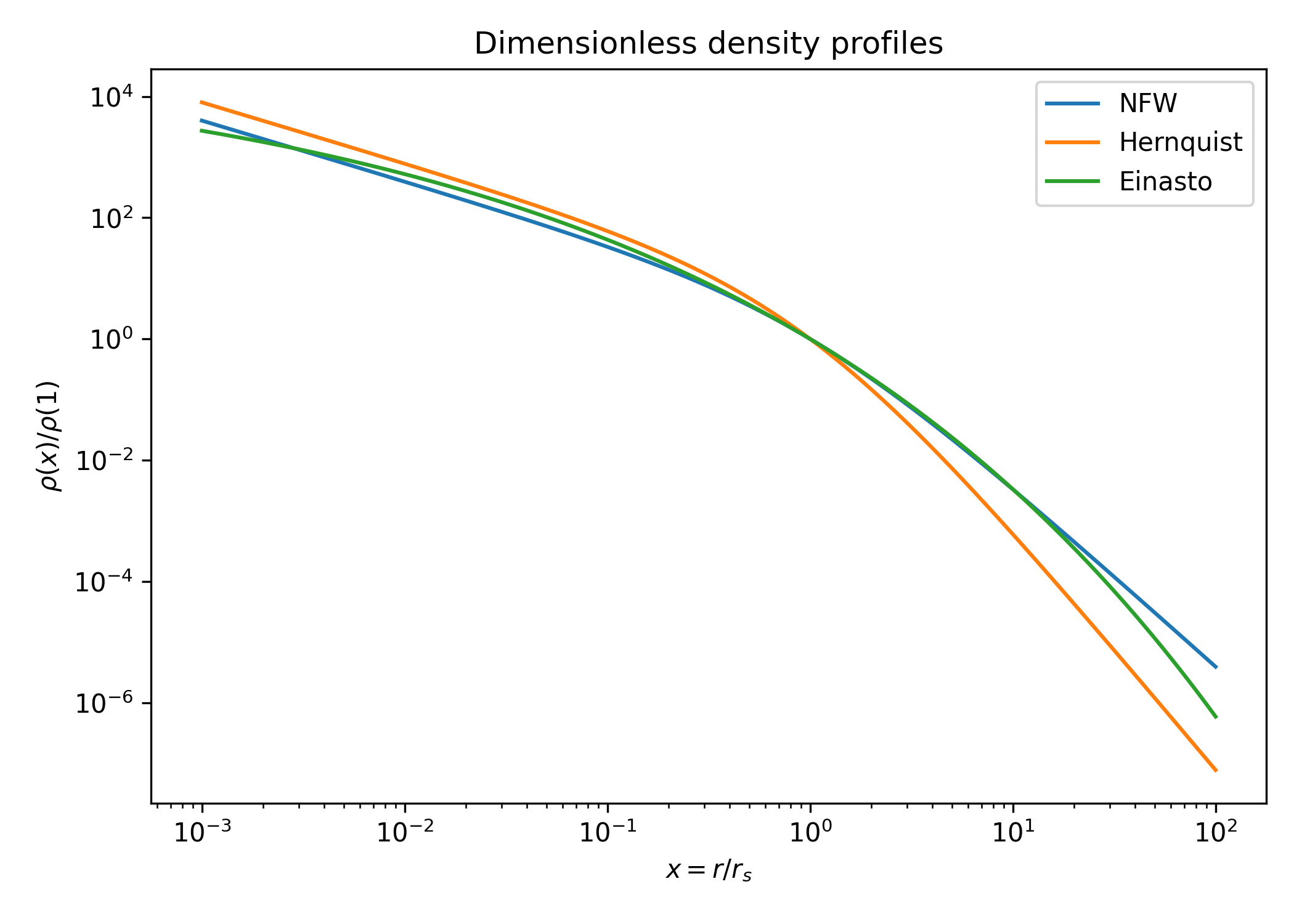}
\includegraphics[width=0.4600\linewidth]{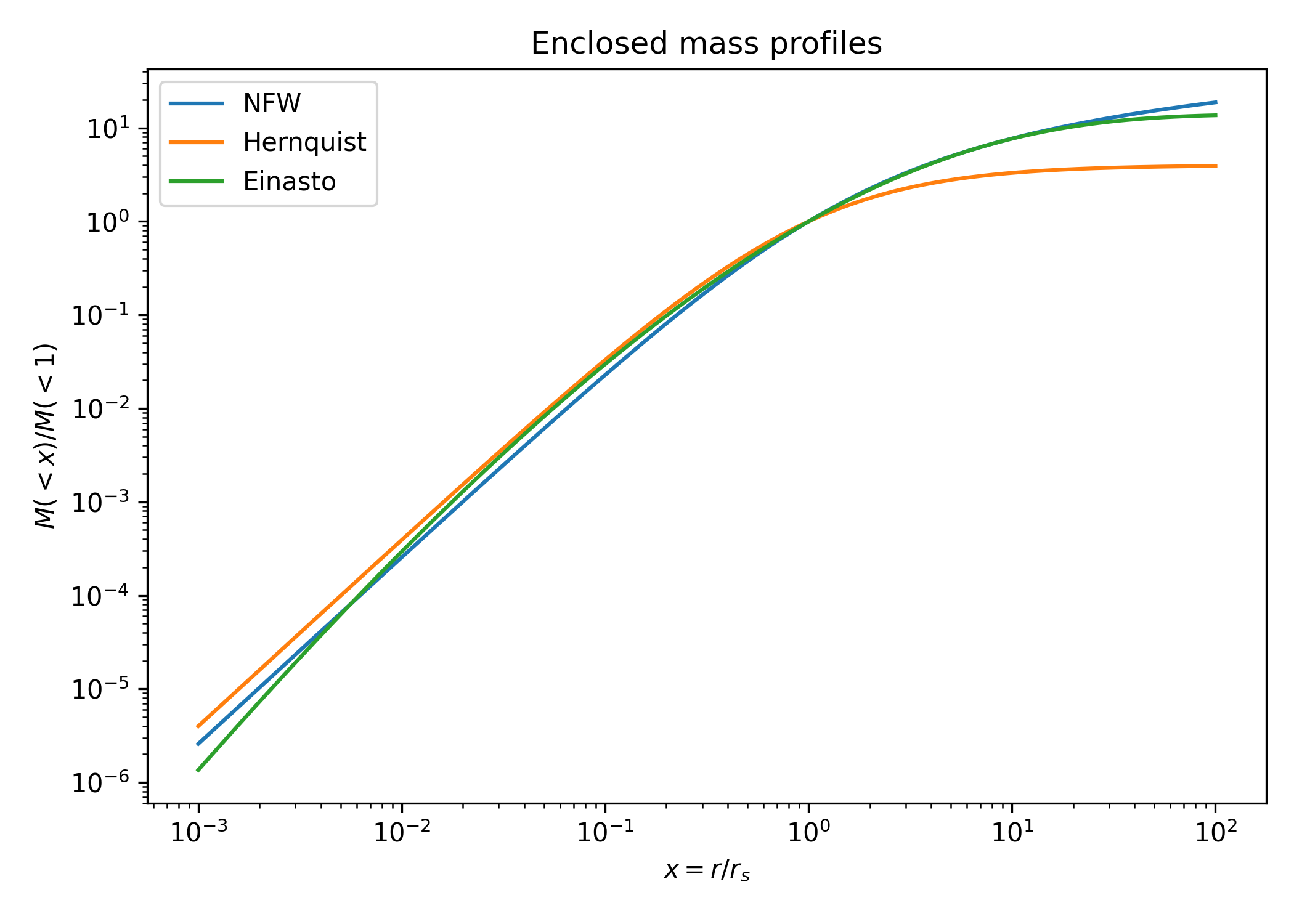}
\includegraphics[width=0.4600\linewidth]{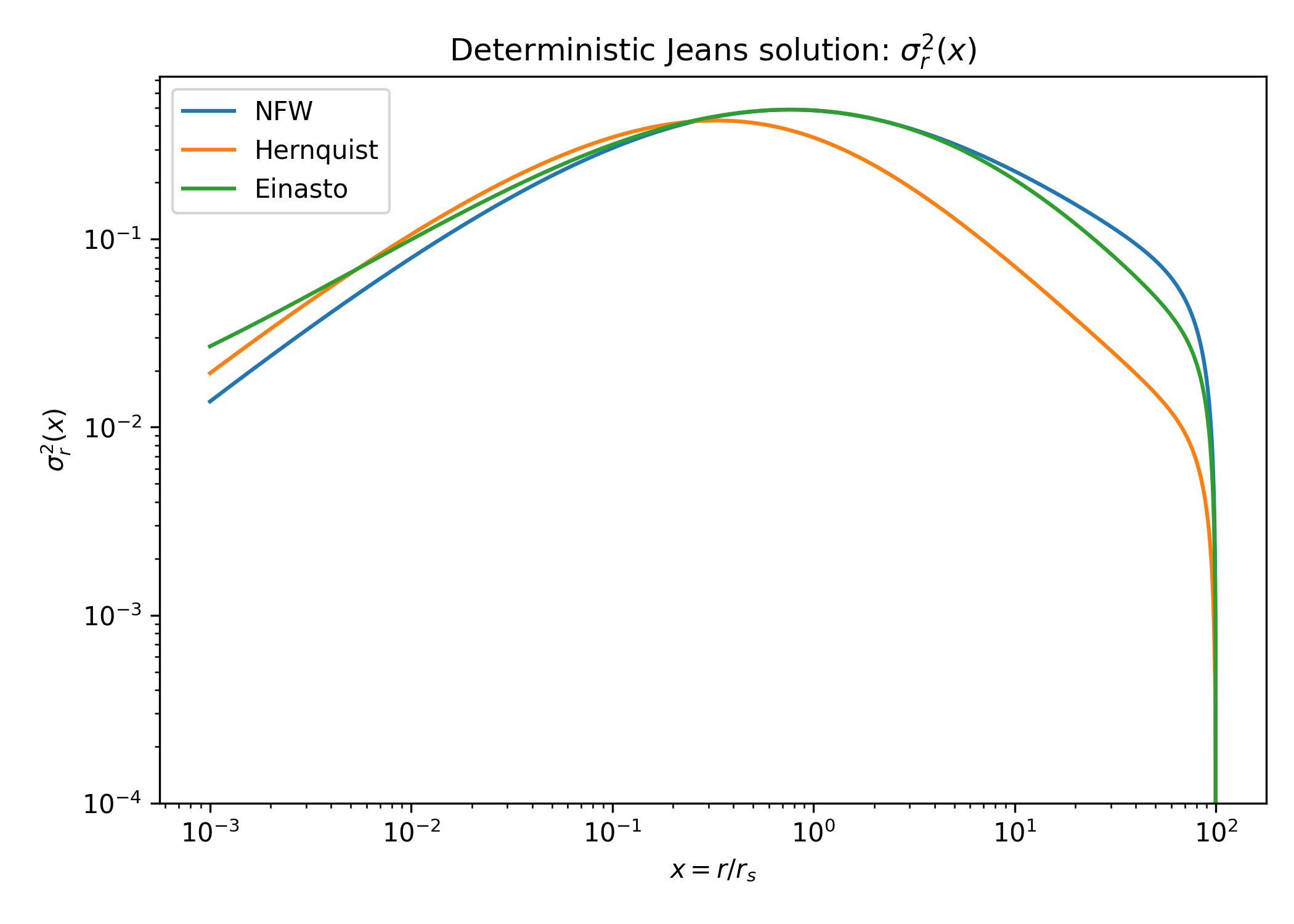}%
\caption{
Dimensionless density, enclosed mass, and deterministic radial
velocity-dispersion profiles for the NFW, Hernquist, and Einasto halo
models. The density and mass panels display the structural differences
between the adopted profiles, while the lower panel shows the
corresponding zero-noise Jeans solution for $\sigma_r^2(x)$ with
$x=r/r_s$. The NFW profile is used as the cosmological benchmark, the
Hernquist model provides a finite-mass analytical comparison, and the
Einasto profile represents a curved-slope phenomenological halo model.
}
\label{fig:density_mass_sigma}
\end{figure}

Before introducing stochastic effects, it is useful to compare the
deterministic ingredients entering the Jeans equation for the three halo
models considered in this work. Figure~\ref{fig:density_mass_sigma}
shows the dimensionless density profiles, the enclosed-mass profiles,
and the corresponding deterministic radial velocity dispersion obtained
from the spherical Jeans equation. The profiles are normalized in a
common way in order to isolate differences in shape rather than overall
mass scale. The NFW profile provides the standard cosmological
benchmark, the Hernquist model has the same inner cusp but a finite
total mass, and the Einasto profile displays a continuously varying
logarithmic slope. These structural differences propagate directly into
the gravitational source term and therefore into the deterministic
velocity-dispersion profile.

\begin{figure}[t]
\centering
\includegraphics[width=0.900\linewidth]{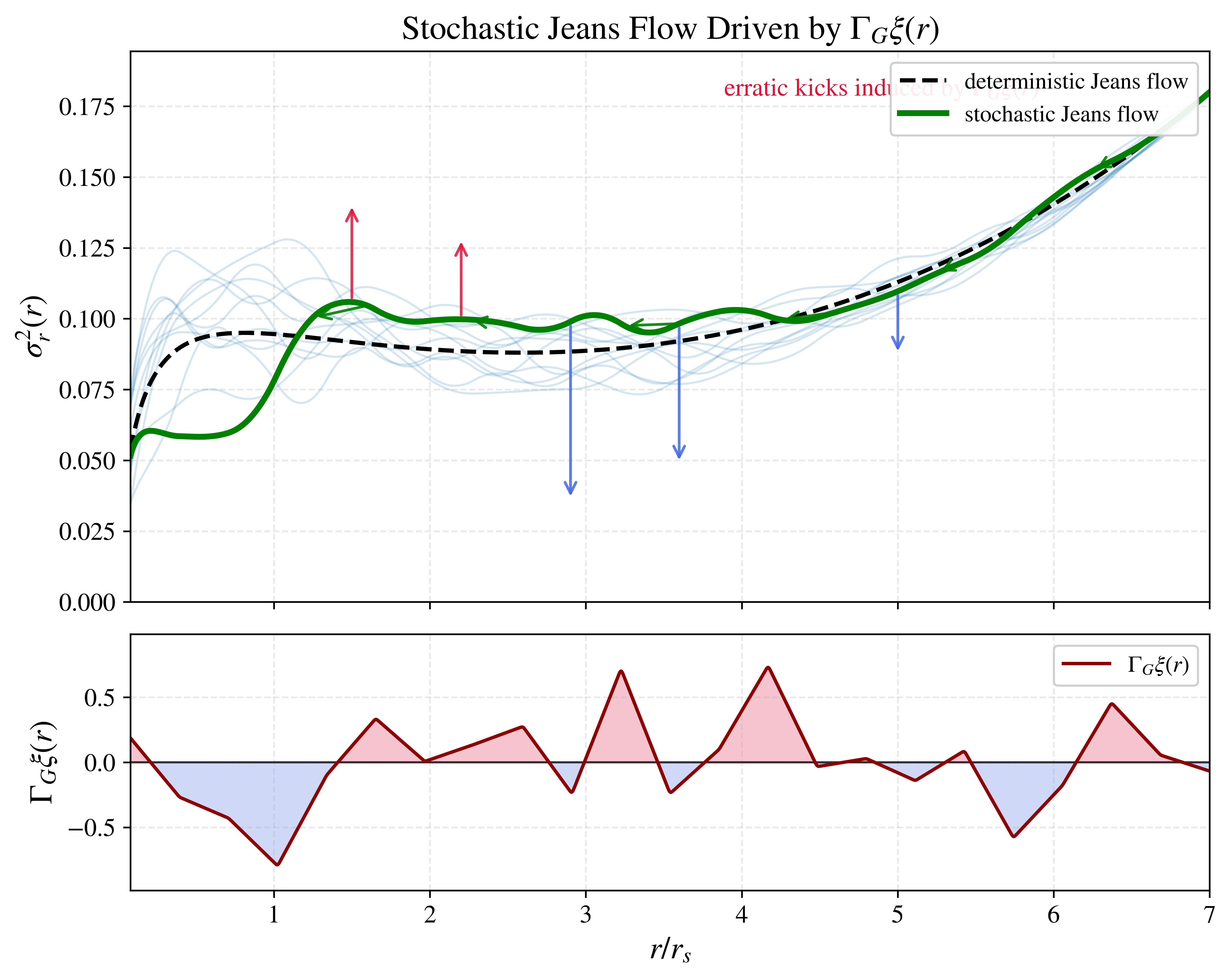}
\caption{
Stochastic Jeans flow driven by a fluctuating gravitational contribution
$\Gamma_G\xi(r)$. The lower panel shows one realization of the stochastic
forcing, with positive and negative regions corresponding to local
enhancements or suppressions of the effective gravitational acceleration.
The upper panel shows the corresponding effect on
$\sigma_r^2(r)$: the black dashed curve is the deterministic Jeans
solution, the thick green curve is one stochastic realization, and the
thin blue curves represent an ensemble of stochastic paths. The vertical
arrows indicate local stochastic kicks induced by $\Gamma_G\xi(r)$. This
figure illustrates how scalar-sector fluctuations are propagated into
halo kinematics through the stochastic Jeans equation.
}
\label{fig:stochastic_jeans_flow}
\end{figure}

Figure~\ref{fig:stochastic_jeans_flow} gives a complementary visualization
of the stochastic forcing itself. The lower panel shows the fluctuating
gravitational contribution $\Gamma_G\xi(r)$, which represents the
coarse-grained effect of spatial fluctuations in the scalar sector on
the effective gravitational coupling. Positive and negative fluctuations
produce local kicks in the radial velocity-dispersion flow. The upper
panel shows how these kicks perturb the deterministic Jeans trajectory.
The black dashed curve is the deterministic solution obtained when
$\Gamma_G=0$, whereas the green curve is one stochastic realization. The
thin blue curves represent an ensemble of realizations, illustrating how
the stochastic gravitational field broadens the deterministic prediction
without necessarily shifting its mean.

\begin{figure}[t]
\centering
\includegraphics[width=0.4600\linewidth]{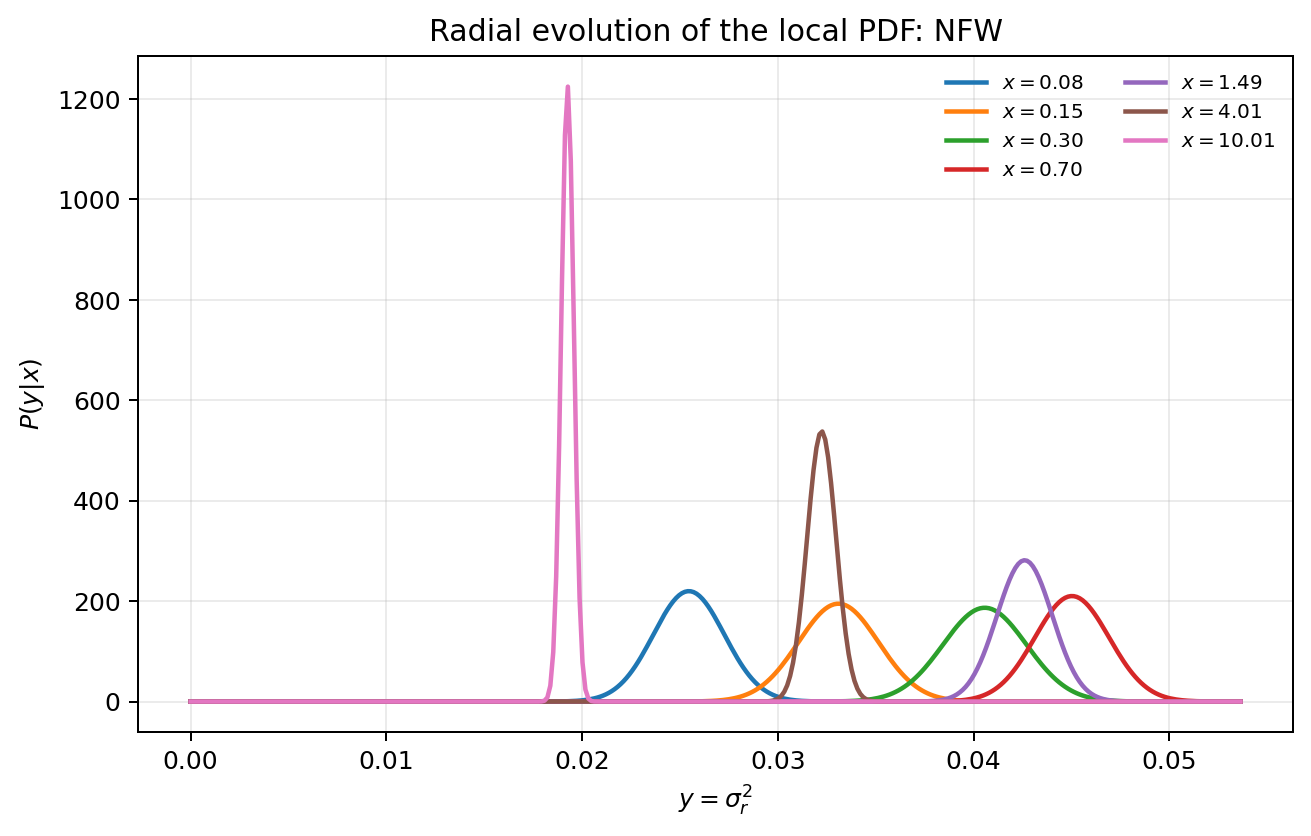}
\includegraphics[width=0.4600\linewidth]{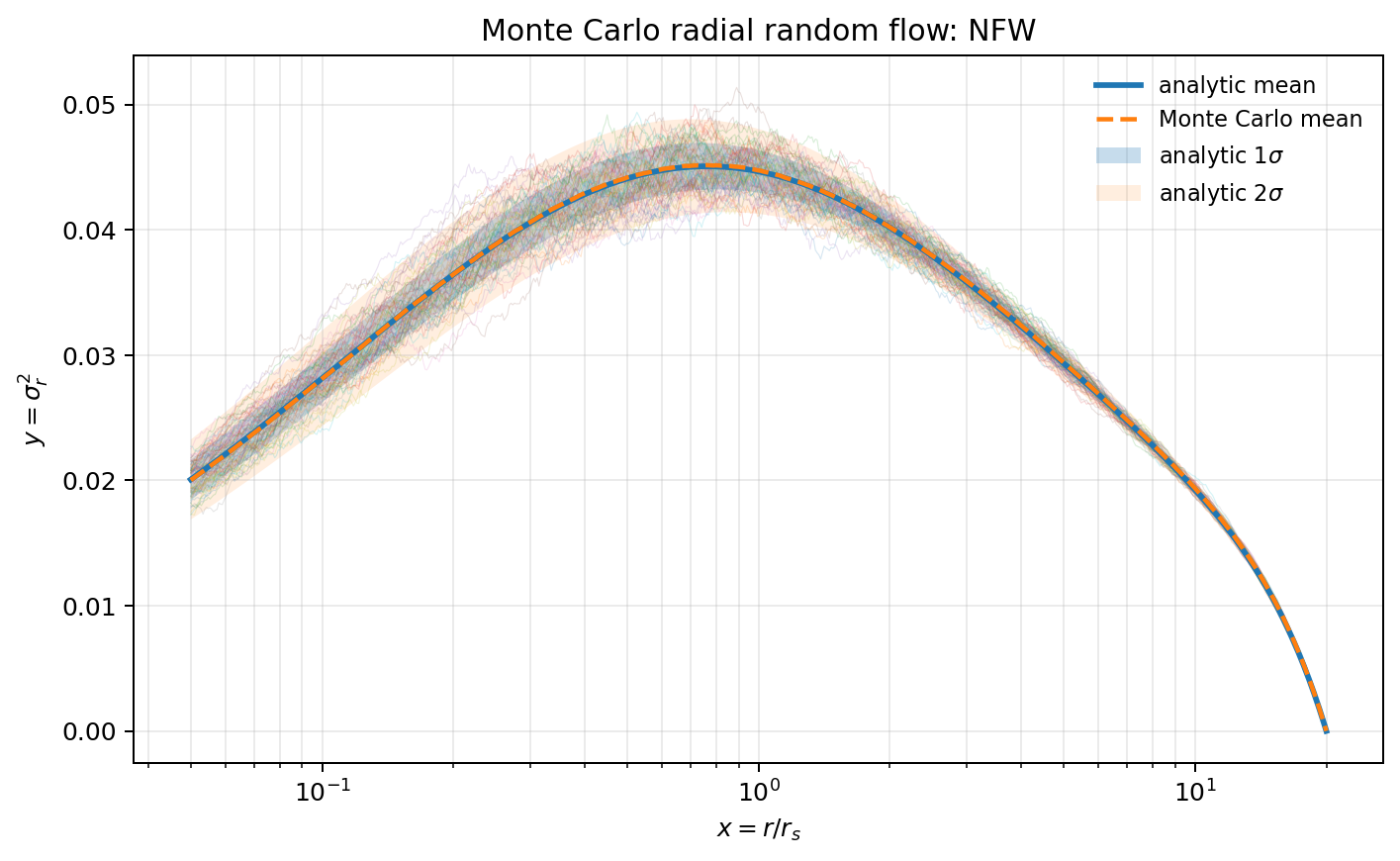}
\includegraphics[width=0.4600\linewidth]{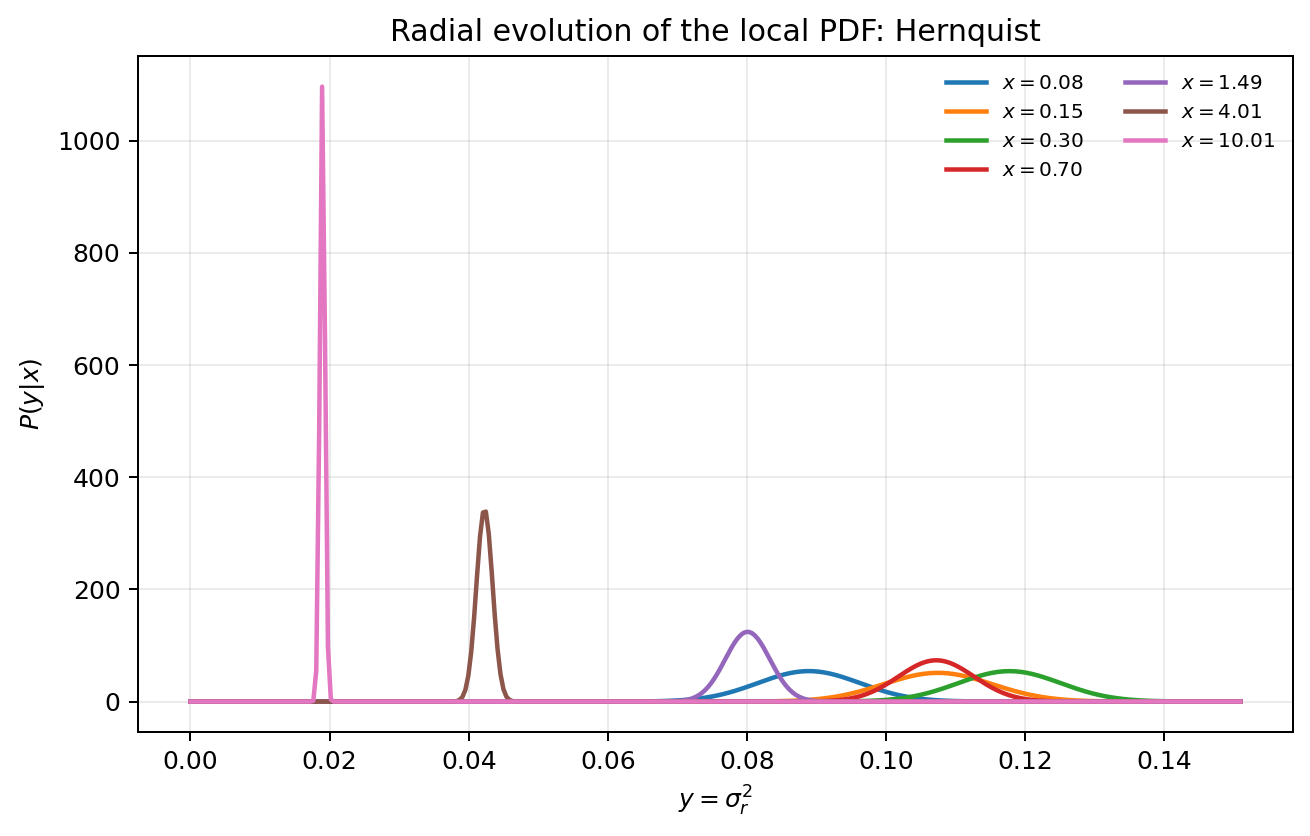}
\includegraphics[width=0.4600\linewidth]{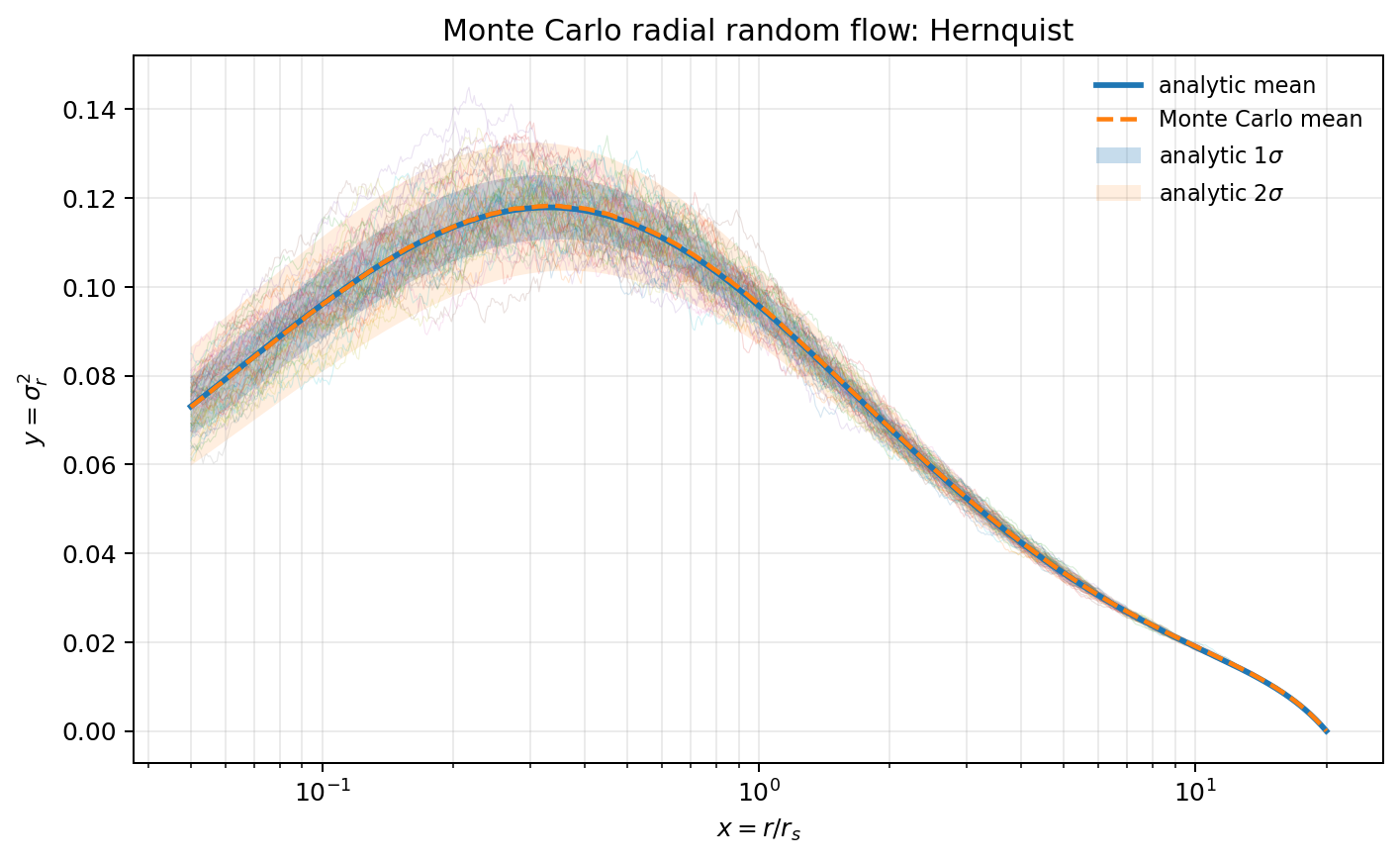}
\includegraphics[width=0.4600\linewidth]{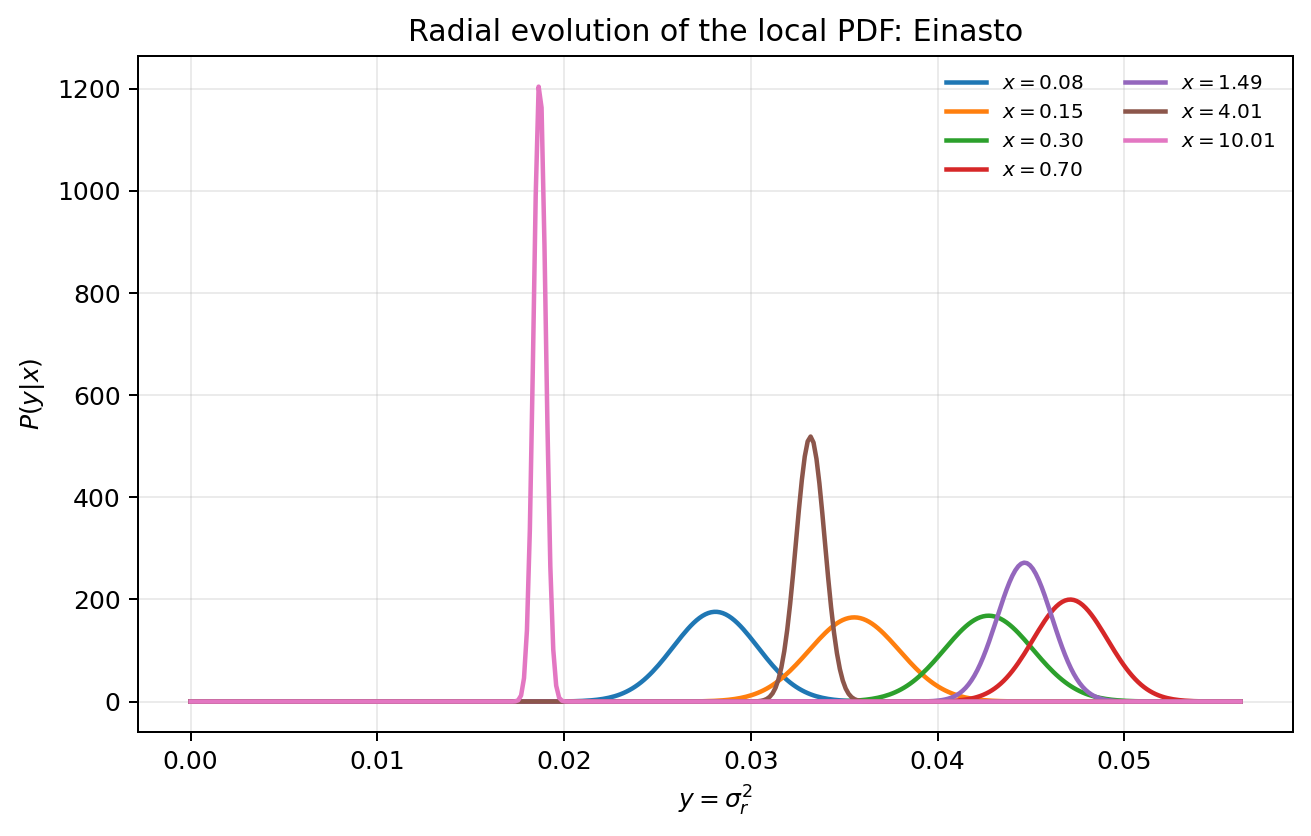}
\includegraphics[width=0.4600\linewidth]{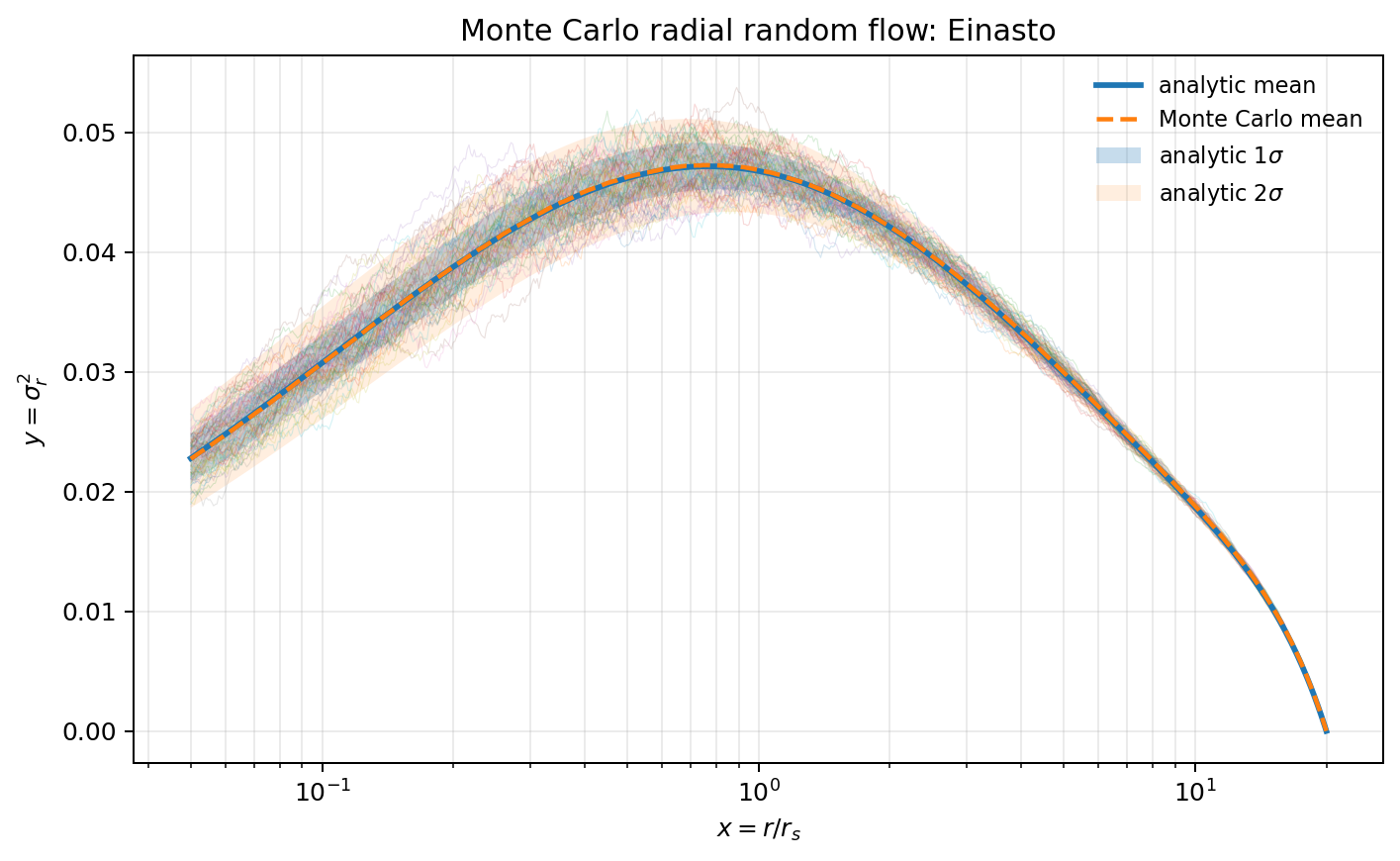}

\caption{
Radial probability distribution and Monte Carlo realization of the
stochastic Jeans flow for the NFW, Hernquist, and Einasto profiles. Left
panels: local probability densities $P(y|x)$ for
$y=\sigma_r^2$ at selected radii. The displacement and narrowing or
broadening of the distributions reflect the radial dependence of the
mean and variance obtained from the Fokker--Planck equation. Right
panels: Monte Carlo realizations of the radial random flow compared with
the analytical mean and the $1\sigma$ and $2\sigma$ probability bands.
The stochastic mean coincides with the deterministic Jeans solution,
while the variance quantifies the broadening induced by fluctuations in
the effective gravitational coupling.
}
\label{fig:radial_pdf_mc}
\end{figure}

The stochastic sector turns the deterministic Jeans solution into a
radial probability distribution. Figure~\ref{fig:radial_pdf_mc} shows
two complementary diagnostics of this construction. The left panels show
local probability densities $P(y|x)$ at different radii, where
$y=\sigma_r^2$. Since the noise is additive and Gaussian, each local
distribution is Gaussian, with mean and variance determined by the
Fokker--Planck solution. The right panels compare the analytical mean
and probability bands with Monte Carlo realizations of the radial random
flow. The agreement between the Monte Carlo ensemble and the analytical
bands verifies that the stochastic broadening is correctly captured by
the moment equations derived from the It\^o formulation.

\begin{figure}[t]
\centering
\includegraphics[width=0.900\linewidth]{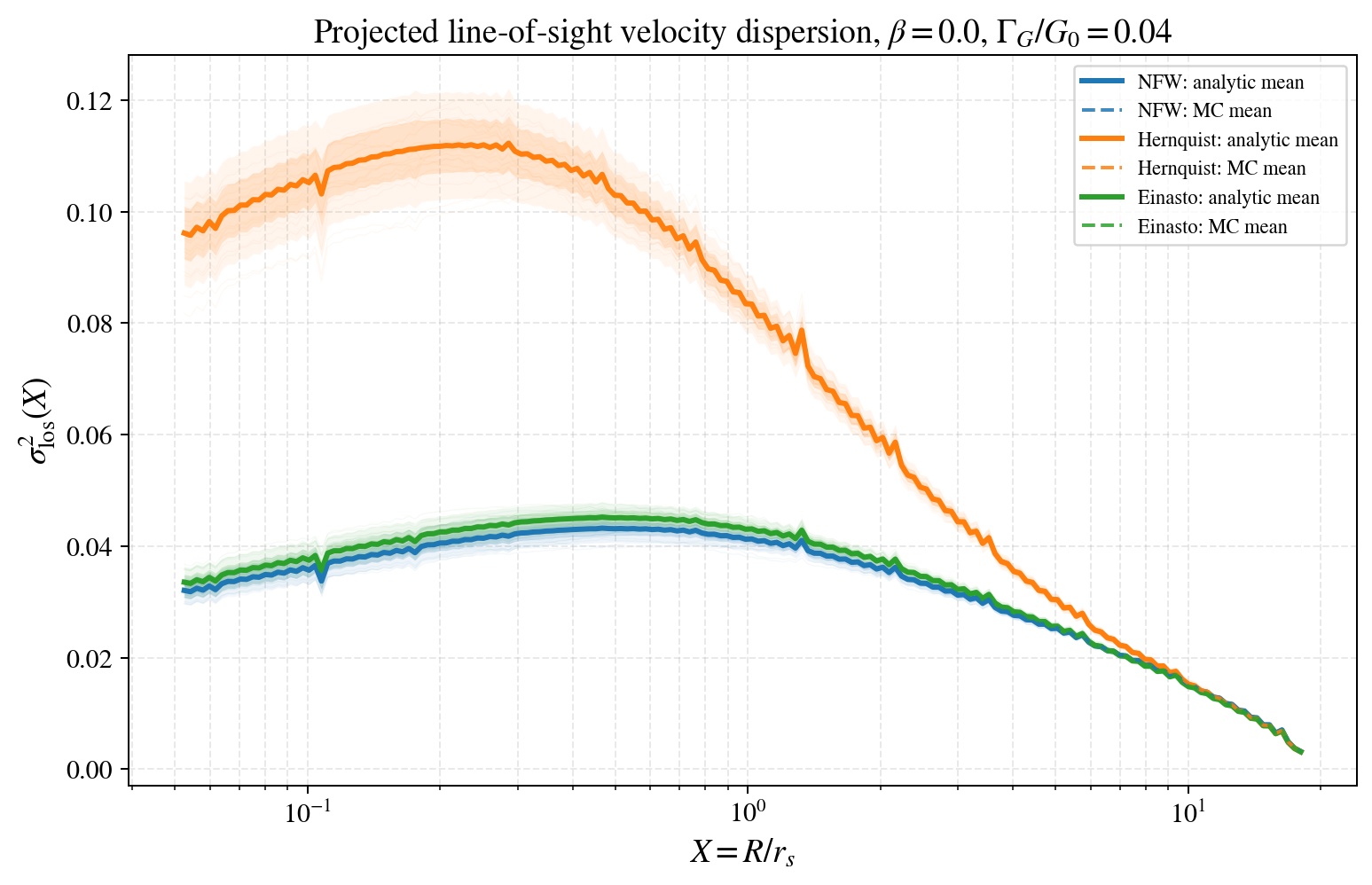}
\caption{
Projected line-of-sight velocity dispersion for the NFW, Hernquist, and
Einasto profiles. Solid curves show the analytical projected mean, shaded
regions show the stochastic uncertainty bands, and dashed curves show the
Monte Carlo projected means. The bands are obtained by propagating the
radial covariance of $\sigma_r^2$ through the line-of-sight projection
kernel.
}
\label{fig:projected_los}
\end{figure}

The projected line-of-sight velocity dispersion provides the most direct
connection between the stochastic Jeans construction and observable halo
kinematics. Figure~\ref{fig:projected_los} shows
$\sigma_{\rm los}^2(R)$ for the three halo profiles, including the
stochastic uncertainty bands. The projected mean is obtained by inserting
the radial mean profile $\mu(r)$ into the standard projection kernel,
whereas the projected variance is computed by propagating the radial
covariance through the same kernel. In this way the figure illustrates
how fluctuations of the scalar gravitational sector generate observable
uncertainty bands in projected kinematic quantities.

A useful way to visualize the stochastic Jeans equation is to regard it
as a radial random flow driven from an outer boundary toward smaller
radii. Since the coefficients of the stochastic equation depend explicitly
on radius, the system is nonautonomous and one should not interpret the
limiting probability distribution as a stationary invariant measure in
the usual autonomous sense. Instead, the relevant object is a pullback
distribution, or equivalently a random attracting graph determined by the
outer boundary condition and by the realization of the stochastic
forcing.

\begin{figure}[t]
\centering
\includegraphics[width=0.900\linewidth]{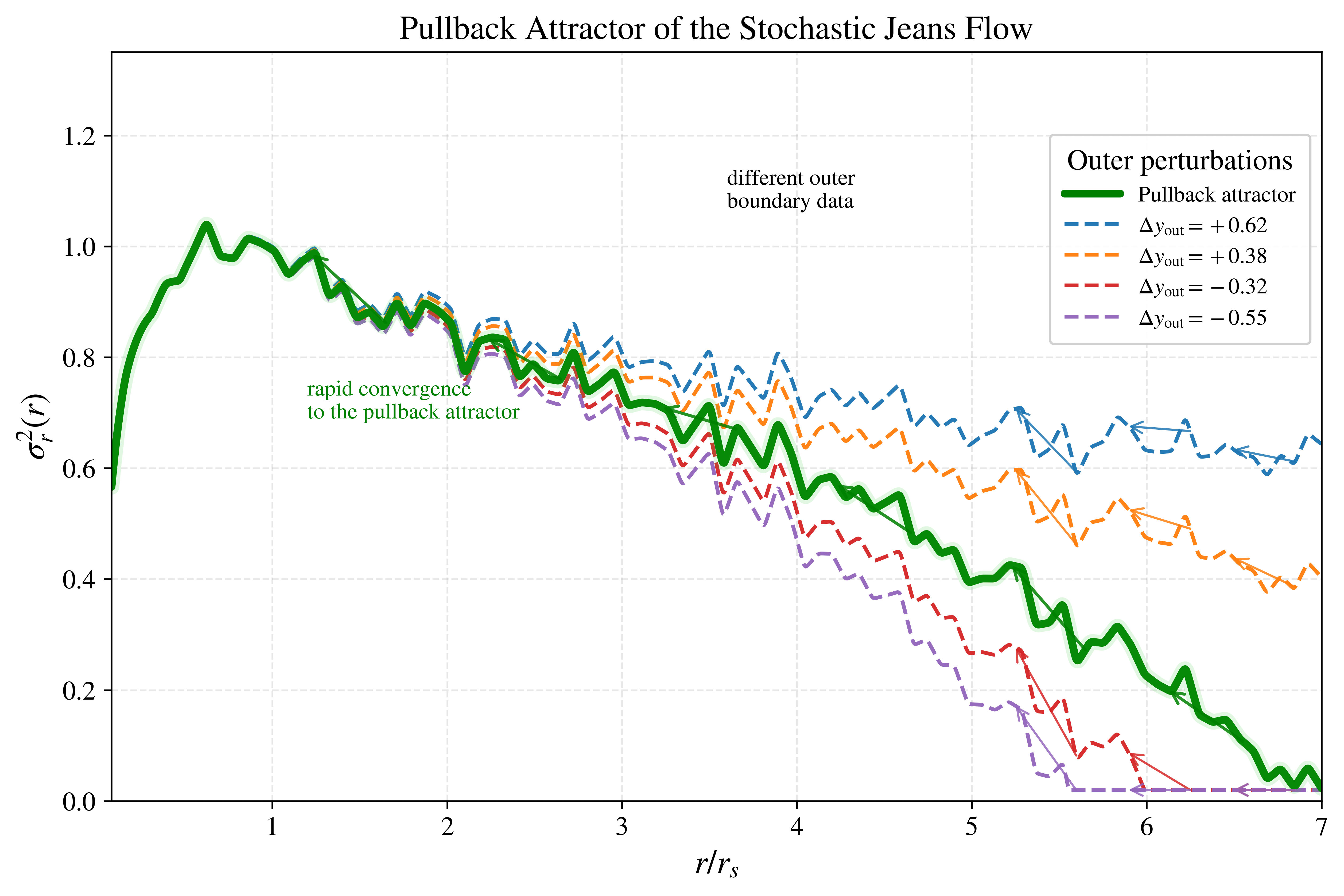}
\caption{
Pullback attractor of the stochastic Jeans flow. The dashed curves
represent solutions initialized at the outer boundary with different
values of $\Delta y_{\rm out}$, while the thick green curve represents
the limiting random profile selected by the stochastic radial flow. The
arrows indicate inward evolution from large to small radii. The
convergence of different boundary-data realizations toward the same
random curve illustrates the pullback character of the solution: the
system loses memory of the imposed outer boundary condition and
approaches a stochastic attracting profile for
$y(r)=\sigma_r^2(r)$. Because the radial flow is nonautonomous, this
object should be interpreted as a pullback attractor or random attracting
graph, not as a stationary fixed point.
}
\label{fig:pullback_attractor}
\end{figure}

Figure~\ref{fig:pullback_attractor} illustrates this idea. Several
trajectories are initialized with different outer boundary values
$\Delta y_{\rm out}$ and then evolved inward under the same stochastic
radial flow. The rapid convergence of these curves toward the thick green
random curve shows the loss of memory of the imposed external boundary
data. In this operational sense, the green curve represents the pullback
attractor of the stochastic Jeans flow: it is the limiting random profile
selected by the dynamics when the boundary data are propagated inward.
This should not be confused with a fixed point in phase space; it is a
random radial profile, or random graph, associated with the
nonautonomous stochastic boundary-value problem.

\subsection{Validity regime and possible extension to infall}
\label{subsec:validity_infall}

The present framework applies to the stationary spherical Jeans regime. It is therefore most appropriate for the quasi-virialized region of halos, where mean radial streaming velocities are negligible compared with random motions. In the outskirts of galaxy clusters, especially between the virial radius, the splashback radius, and the turnaround radius, radial infall becomes dynamically important. In that regime the generalized Jeans equation contains an additional streaming term of the form
\begin{eqnarray}
\bar v_r\frac{\partial \bar v_r}{\partial r}
+
\frac{\partial \bar v_r}{\partial t},
\label{eq:streaming_term}
\end{eqnarray}
where $\bar v_r$ is the mean radial velocity. A stochastic extension of that generalized equation is a natural continuation of the present work, but it should be treated separately because it describes a nonstationary region of halo dynamics.

Taken together, these figures show that the stochastic Jeans equation
provides a controlled way to propagate fluctuations of the effective
gravitational coupling into halo kinematic observables. At the radial
level, the stochastic sector generates a probability distribution for
$\sigma_r^2(r)$ rather than a single deterministic profile. At the
projected level, the same covariance structure produces uncertainty bands
for $\sigma_{\rm los}^2(R)$, which are directly connected with
observational velocity-dispersion measurements. The pullback-attractor
visualization further clarifies the role of the outer boundary condition:
for a fixed stochastic realization, different imposed boundary data may
converge toward the same limiting random profile, showing that the
stochastic radial flow can select an effective attracting solution.

\section{Conclusions}
\label{sec:conclusions}

We have formulated a stochastic extension of the stationary spherical
Jeans equation motivated by scalar--tensor gravity. In this framework,
fluctuations of the scalar sector induce fluctuations in the effective
gravitational coupling, which are represented after coarse graining as
$\Geff(r,\omega)=\Gbar(r)+\Gamma_G(r)\xi(r,\omega)$. This leads to a
linear It\^o stochastic differential equation for
$y(r)=\sigma_r^2(r)$ along the radial coordinate.

The radial coordinate is not a physical time, but it acts as an ordering
parameter for the Jeans boundary-value problem. The stochastic Jeans
equation therefore defines a nonautonomous radial random flow. In the
white-noise limit, the associated Fokker--Planck equation gives the
radial probability distribution $P(y,r)$. Because the noise is additive,
the deterministic Jeans solution is recovered as the mean profile, while
the stochastic sector generates a probability band around it. We derived
closed integral expressions for the variance and covariance of
$\sigma_r^2(r)$ using the It\^o isometry.

The formalism was applied to NFW, Hernquist, and Einasto halo models.
For NFW and Hernquist profiles, the relevant kernels are written in terms
of elementary functions, while the Einasto enclosed mass involves the
lower incomplete gamma function. In all cases, the stochastic mean,
variance, and covariance reduce to one-dimensional quadratures once the
profile, anisotropy, mean coupling, fluctuation amplitude, and outer
boundary are specified.

The numerical examples illustrate how stochastic fluctuations of the
effective gravitational coupling propagate into halo kinematic
observables. At the radial level, the model predicts a probability
distribution for $\sigma_r^2(r)$. At the projected level, the radial
covariance produces uncertainty bands for the line-of-sight velocity
dispersion $\sigma_{\rm los}^2(R)$. Monte Carlo realizations of the
radial random flow agree with the analytical moment equations, providing
a direct consistency check of the Fokker--Planck formulation.

We also emphasized the pullback interpretation of the stochastic Jeans
flow. Since the system is nonautonomous in radius, the limiting random
profile should be understood as a pullback attracting graph selected by
the inward propagation from the outer boundary, rather than as a
stationary invariant measure. This interpretation clarifies how different
outer boundary perturbations can lose memory of their initial values and
approach a common stochastic profile.

The present work is restricted to the stationary spherical Jeans regime.
Applications to cluster outskirts, splashback, or turnaround regions
require a generalized Jeans equation with mean radial streaming and
possible explicit time dependence. These extensions will be important for
connecting the present stochastic framework with nonstationary halo
dynamics and will be pursued in future work.

\section*{Acknowledgments}

I would like to thank Maria Margarita and Miguel Amado for their continuous inspiration and support, and the Foundation for Research Support of Espírito Santo (FAPES) for the partial support for the present work.


\end{document}